\documentclass[final,5p,times,twocolumn]{elsarticle}

\usepackage{amssymb}

\usepackage{amsmath}
\usepackage{xcolor}

\usepackage[utf8]{inputenc}

\journal{Physics Letters B}

\begin{document}

\begin{frontmatter}



\title{Double D-meson production in proton-proton and proton-lead collisions at the LHC}


\author{Ilkka Helenius}
\ead{ilkka.m.helenius@jyu.fi}

\author{Hannu Paukkunen}
\ead{hannu.paukkunen@jyu.fi}

\address{University of Jyvaskyla, Department of Physics, P.O. Box 35, FI-40014 University of Jyvaskyla, Finland}
\address{Helsinki Institute of Physics, P.O. Box 64, FI-00014 University of Helsinki, Finland}

\begin{abstract}
We consider the simultaneous production of two heavy-flavoured hadrons -- particularly D mesons -- at the LHC. We base our calculations on collinearly factorized QCD at next-to-leading order, using the contemporary parton distribution functions and D-meson fragmentation functions. The contributions of double-parton scatterings are included in the approximation of independent partonic interactions. Our framework benchmarks well with the available proton-proton data from the LHCb collaboration giving us confidence to make predictions for proton-lead collisions. Our results indicate that the double D-meson production in proton-lead collisions should be measurable at the LHCb kinematics with the already collected Run-II data, and should provide evidence for double-parton scattering at perturbative scales with a nuclear target.
\end{abstract}


\end{frontmatter}


\section{Introduction}
\label{Introduction}

The recent measurements of inclusive open heavy-flavour -- particularly D and B mesons -- in proton-proton (p-p) collisions at the CERN Large Hadron Collider (LHC) \cite{Aaij:2013mga,Aaij:2015bpa,Aaij:2016jht,Aaij:2017qml,Acharya:2017jgo,Acharya:2019mgn,Aad:2015zix,Sirunyan:2017xss} provide opportunities to expose different facets of Quantum Chromodynamics (QCD) \cite{Andronic:2015wma}. On one hand, due to the heavy-quark mass which serves as a hard interaction scale, the perturbative QCD calculations \cite{Nason:1989zy,Mangano:1991jk,Cacciari:1998it,Frixione:2003ei,Kniehl:2004fy,Helenius:2018uul} can be extended e.g. to very small transverse momenta ($p_{\rm T}$) where the calculations with massless quarks become inherently invalid. As the measurements at low $p_{\rm T}$ are statistically very precise, they offer an ideal testbed to benchmark perturbative calculations at low interaction scales. On the other hand, open heavy-flavour production can be used as a tool to probe non-perturbative aspects of heavy-quark fragmentation \cite{Anderle:2017cgl} and the quark-gluon structure of protons and nuclei \cite{Eskola:2019bgf,Zenaiev:2015rfa,Gauld:2015yia,Gauld:2016kpd,Gauld:2015lxa,Lansberg:2016deg,Kusina:2017gkz,Cacciari:2015fta}. The low-$p_{\rm T}$ open heavy-flavour production in proton-lead (p-Pb) collisions \cite{Adam:2016ich,Aaij:2017gcy,Aaij:2019lkm,Aaij:2018ogq} may also open prospects to disentangle non-linear saturation \cite{Fujii:2013yja,Ducloue:2016ywt} and collinearly factorized QCD pictures in a regime where both should be valid descriptions.

The inclusive production of two D mesons provides exciting further opportunities. While the heavy-quarks are predominantly produced in pairs, the experimental overall reconstruction efficiencies for two D-meson final states are low and, roughly, only one out of million primarily produced double-D events can be reconstructed. Nevertheless, simultanous production of two D mesons has been observed in p-p \cite{Aaij:2012dz} and p-$\overline{\rm p}$ \cite{Reisert:2007zza} collisions. This offers possibilities to test e.g. the heavy-quark vs. heavy-antiquark asymmetries \cite{Gauld:2015qha} and, in particular, to study the double-parton scattering (DPS) \cite{Belyaev:2017sws,Blok:2016lmd,Cazaroto:2016nmu,Maciula:2016wci,vanHameren:2014ava}. While the formal theory of factorization in DPS has recently advanced significantly \cite{Diehl:2017wew,Diehl:2017kgu,Bartalini:2017jkk}, we still know relatively little of the non-perturbative structure of e.g. the double-parton distributions (dPDFs) \cite{Gaunt:2009re} which would be required in precise phenomenological applications. Thus, simplifying assumptions concerning DPS have to be made which can lead to apparent shortcomings. For instance, measurements are often interpreted in terms of an \emph{effective cross section} $\sigma_{\rm eff}$ whose inverse is proportional to the DPS probability. Its value has been observed to differ significantly depending from which observable it is extracted \cite{Aaboud:2018tiq}. It is conceivable that this is due to overly simplifying the problem of DPS or, alternatively, overlooking the contributions of single-parton scattering (SPS) \cite{Karpishkov:2019vyt}. A complementary approach to hard DPS is provided by Monte-Carlo event generators in which the soft or semi-hard multiparton interactions (MPIs) give rise to the underlying event found necessary to describe the multiplicity distributions in hadronic high-energy collisions \cite{Bartalini:2017jkk, Bartalini:2011jp}.

The generic prediction is that in proton-nucleus (p-$A$) collisions the DPS signal gets enhanced in comparison to p-p case, due to the possibility of proton to interact with two or more nucleons simultaneously \cite{Strikman:2001gz,dEnterria:2012jam,dEnterria:2013mrp,dEnterria:2017yhd,Huayra:2019iun}. As in p-p collisions, multiple interactions are necessary to explain the multiplicity distributions in collisions involving heavy nuclei \cite{Deng:2010mv, Bierlich:2018xfw}, but a clean experimental confirmation for hard DPS processes is still lacking. As we will conclude later in this letter, it seems realistic that the double D-meson production could provide the first direct evidence of DPS in p-Pb collisions at clearly perturbative scales, and that the signal should be visible already in the collected Run-II data. In reaching this conclusion we have first confronted our QCD framework with the LHCb p-p data and a reasonble agreement there encourages us to apply it to p-Pb collisions. Before getting into the actual results we will first describe our theoretical framework in the next two sections. 

\section{Double-inclusive production in nuclear collisions}
\label{DoubleInclusive}

We will estimate the double-inclusive cross sections in collision of two nuclei, $A$ and $B$, in terms of inclusive per-nucleon SPS cross sections $\sigma^{\rm sps}_{\rm nn \rightarrow \mathcal{O} + X}$ as
\begin{align}
& \frac{\mathrm{d}\sigma_{AB \rightarrow a + b + X}}{\mathrm{d}^3\vec{p}^{\, a} \mathrm{d}^3\vec{p}^{\, b}} = 
AB  \left[ \frac{\mathrm{d}\sigma^{\rm sps}_{nn \rightarrow a + b + X}}{\mathrm{d}^3\vec{p}^{\, a} \mathrm{d}^3\vec{p}^{\, b}}  +  
\frac{m}{\sigma_{\rm eff}^{AB}} \frac{\mathrm{d}\sigma^{\rm sps}_{nn \rightarrow a + X}}{\mathrm{d}^3 \vec{p}^{\,a} } \frac{\mathrm{d}\sigma^{\rm sps}_{nn \rightarrow b + X}}{\mathrm{d}^3\vec{p}^{\,b} } \right]  
\label{eq:finaldouble} 
\end{align}
where $\vec{p}^{\, a}$ and $\vec{p}^{\, b}$ refer to the momenta of the produced particles $a$ and $b$. If $a$ and $b$ are identical particles $m=1/2$, and $m=1$ otherwise. In the case of independent partonic interactions, the effective cross section $\sigma_{\rm eff}^{AB}$ in $A$-$B$ collision is process independent and can be interpreted as a purely geometric object, 
\begin{align}
\frac{1}{\sigma_{\rm eff}^{AB}} \equiv \Bigg\{ \frac{ 1}{\sigma_{\rm eff}}  
&  + \frac{(B-1)}{B^2} \int \mathrm{d}^2\vec{B} \left[ T_{nB}\left( \vec{B}\right) \right]^2  \label{eq:sigmaeffAB} \\
&  + \frac{(A-1)}{A^2} \int \mathrm{d}^2\vec{B} \left[ T_{nA}\left( \vec{B}\right) \right]^2 \nonumber \\
&  + \frac{(A-1)(B-1)}{(AB)^2} \int \mathrm{d}^2\vec{B} \left[ T_{AB}\left( \vec{B}\right) \right]^2 \Bigg\} \nonumber \,.
\end{align}
Here, 
\begin{equation}
 \frac{1}{\sigma_{\rm eff}} = \int \mathrm{d}^2\vec{b} \left[ t_{\rm nn}\left( \vec{b}\right) \right]^2 \,,
\end{equation}
where $t_{\rm nn}(\vec{b})$ is the overlap function between two nucleons at fixed impact parameter $\vec{b}$. In geometric sense, we would write
\begin{equation}
t_{\rm nn}(\vec {b}) \equiv \int_{-\infty}^{\infty} \mathrm{d}^2{\vec s} \, t_{\rm n}({\vec s} + {\vec b}/2)t_{\rm n}({\vec s} - {\vec b}/2) \,,
\end{equation}
where $t_{\rm n}({\vec s})$ is the transverse profile of nucleons obtained by integrating the density of nucleons $\rho^{\rm n}$ over the longitudinal spatial component,
\begin{equation}
t_{\rm n}({\vec s}) \equiv \int_{-\infty}^{\infty} \mathrm{d}z\, \rho^{\rm n}(\vec{s},z) \,.
\end{equation}
The overlap functions $T_{\mathrm{n}A}(\vec{B})$ and $T_{AB}(\vec{B})$ at fixed impact parameter $\vec{B}$ are here defined as \cite{Florkowski:2010zz}
\begin{align}
T_{\mathrm{n}A}(\vec{B}) & \equiv \int_{-\infty}^{\infty} \mathrm{d}^2{\vec s} \, t_{\rm nn}({\vec s} + {\vec B}/2)T_{A}({\vec s} - {\vec B}/2) \\
& \approx T_{A}({\vec B}) \,, \nonumber
\end{align}
where the approximation holds for point-like nucleons, and
\begin{align}
T_{AB}(\vec{B}) & \equiv \int_{-\infty}^{\infty} \mathrm{d}^2{\vec s^{\, A}} \mathrm{d}^2{\vec s^{\, B}} \, 
T_{A}(\vec s^{\, A})
T_{B}(\vec s^{\, B})\,
t_{\rm nn}({\vec B} + {\vec s^B} - {\vec s^A}) \nonumber \\
& \approx
\int_{-\infty}^{\infty} \mathrm{d}^2{\vec s} \, 
T_{A}(\vec s + {\vec B}/2)
T_{B}(\vec s - {\vec B}/2) \,,
\end{align}
where $T_{A}(\vec{S})$ is the standard nuclear thickness function 
\begin{equation}
T_{A}(\vec{s}) \equiv \int_{-\infty}^{\infty} \mathrm{d}z\, \rho^{A}(\vec{s},z) \,,
\end{equation}
and $\rho^{A}$ denotes the density of nuclei. In our notation, the normalization is
\begin{equation}
\int \mathrm{d}^2\vec{s} \, T_{A}(\vec{s}\,) = A. \label{eq:norm}
\end{equation}

Typically, the DPS contribution in Eq.~(\ref{eq:finaldouble}) is derived \cite{dEnterria:2012jam,dEnterria:2013mrp,dEnterria:2017yhd, Diehl:2011yj} by writing the DPS cross section in terms of dPDFs,  and assuming that the dPDFs factorize into a product of single-parton PDFs and that the partonic cross sections for the two subprocesses are unrelated and spatially independent. Alternatively, Eq.~(\ref{eq:finaldouble}) can be derived from an eikonal model for multiparton interactions. Indeed, in a Glauber-type approach, the total cross section for a collision of nuclei A and B can be written in the impact-parameter space as
\begin{equation}
 \sigma_{AB}^{\rm total} = \int \mathrm{d}^2\vec{B} \sum_{k=1}^{AB} \mathcal{P}_k  ( \vec B ) \label{eq:totAB} \,,
\end{equation}
where $\mathcal{P}_k  ( \vec B )$ is the probability of exactly $k$ nucleon-nucleon interactions at fixed impact parameter $ \vec B $, 
\begin{align}
 \mathcal{P}_k  ( \vec B ) & = \int \, 
\left[ \prod_{i=1}^{A} \mathrm{d}\vec{S}^{A}_i \frac{T_{A}\left(\vec{S}^{A}_i \right)}{A} \right]
\left[ \prod_{i=1}^{B} \mathrm{d}\vec{S}^{B}_i \frac{T_{B} \left(\vec{S}^{B}_i \right)}{B} \right] \label{eq:tot} \\
&\hspace{-2.8em} \times \hspace{-0.3em} \sum_{\alpha_{11}=0}^1 \hspace{-0.4em} \ldots \hspace{-0.4em} \sum_{\alpha_{AB}=0}^1
\left[ p_{11}({\alpha_{11}}) \, p_{12}({\alpha_{12}}) \, \cdots \, p_{AB}({\alpha_{AB}}) \right] \delta_{k, \alpha_{11}+\ldots+\alpha_{AB}}
\nonumber \,.
\end{align}
In this expression, we have defined 
\begin{align}
p_{ij}({\alpha_{ij}}) \equiv 
\left( t_{ij} \, \sigma_{\rm nn}^{\rm total} \right)^{\alpha_{ij}}
\left(1 - t_{ij} \, \sigma_{\rm nn}^{\rm total} \right)^{1-\alpha_{ij}} 
\end{align}
where $t_{ij}$ is an abbreviation for the overlap function between two nucleons
\begin{equation}
t_{ij} \equiv t_{\rm nn}\left( \vec B + \vec{S}^{A}_i - \vec{S}^{B}_j \right) \,.
\end{equation}
The second line in Eq.~(\ref{eq:tot}) thus corresponds to the probability of getting exactly $k$ nucleon-nucleon interactions (and $AB-k$ missing ones) at fixed geometric configuration. The total cross section in a single nucleon-nucleon collisions $\sigma_{\rm nn}^{\rm total}$ is given by
\begin{equation}
\sigma_{\rm nn}^{\rm total} = \int {\mathrm{d}^2}\vec{b} \sum_{k=1}^\infty p_k (\vec{b}) \,,
\end{equation}
where the probability $p_k (\vec{b})$ for $k$ partonic interactions is considered to be Poissonian,
\begin{equation}
p_k (\vec{b}) = \exp \left[ -t_{\rm nn}(\vec{b}) \sigma_{\rm nn} \right] \frac{\left[t_{\rm nn}(\vec{b}) \sigma_{\rm nn}\right]^k}{k!} \,. \label{eq:poisson}
\end{equation}
The quantity $\sigma_{\rm nn}$ appearing in Eq.~(\ref{eq:poisson}) is the integrated inclusive cross sections,
\begin{align}
\sigma_{\rm nn} & = \sum_f \int \mathrm{d}{\rm PS}_f \frac{d\sigma^{\rm sps}_{\rm nn \rightarrow f}}{\mathrm{d}{\rm PS}_f} \,, \ \ \mathrm{d}{\rm PS}_f = \prod_{i \in f} \mathrm{d}^3\vec{p}_i 
\end{align}
where the summation is over all \emph{exclusive} final states $f$. We will always make a distinction between the (intensive) total cross section like $\sigma_{\rm nn}^{\rm total}$, and (extensive) integrated cross section like $\sigma_{\rm nn}$. The double-inclusive cross section can now be written as
\begin{align}
& \frac{\mathrm{d}\sigma_{AB\rightarrow a + b + X}}{\mathrm{d}\vec p^{\, a} \mathrm{d}\vec p^{\, b}} = \int \mathrm{d}\vec{B} \sum_{k=1}^{AB} \mathcal{P}_k  ( \vec B ) \label{eq:masterdoublydiff} \\
& 
\prod_{r=1}^k
\int {\mathrm{d}^2}\vec{b}_r \sum_{k_r=1}^\infty \frac{p_{k_r} (\vec{b}_r) }{\sigma_{\rm nn}^{\rm total}}
\prod_{\ell=1}^{k_r} 
\left[\frac{1}{\sigma_{\rm nn}} \sum_{f_{r\ell}} \int \mathrm{d}{\rm PS}_{f_{r\ell}} \frac{\mathrm{d}\sigma^{\rm sps}_{nn \rightarrow f_{r\ell}}}{\mathrm{d}{\rm PS}_{f_{r\ell}}} \right] 
\nonumber \\
& \left[ \sum_{i=1}^{k} \sum_{j=1}^{k_r} \sum_n \delta^{\,(3)} \left(\vec p^{\, a} - \vec p_{ij}^{\, a_n} \right)
\times \left\{
\begin{array}{c}
 1, \ {\rm if} \ a_n \in f_{ij}\\
 0, \ {\rm if} \ a_n \notin f_{ij}
\end{array}
\right\}
\right] \nonumber \\
& \left[ \sum_{i=1}^{k} \sum_{j=1}^{k_r} \sum_n \delta^{\,(3)} \left(\vec p^{\, b} - \vec p_{ij}^{\, b_n} \right)
\times \left\{
\begin{array}{c}
 1, \ {\rm if} \ b_n \in f_{ij}\\
 0, \ {\rm if} \ b_n \notin f_{ij}
\end{array}
\right\}
\right] \,. \nonumber 
\end{align}
In the equation above, each term is a product of the form
\begin{equation}
{\mathcal{P}_k ( \vec B )} \ \times \ \left[ \prod_{r=1}^k {p_{k_r} (\vec{b}_r)} \right] \ \times \ \left[ \prod_{\ell=1}^{k_r} \frac{1}{\sigma_{\rm nn}} \frac{\mathrm{d}\sigma^{\rm sps}_{nn \rightarrow f_{r\ell}}}{\mathrm{d}{\rm PS}_{f_{r\ell}}} \right] \,,
\end{equation}
corresponding to the total probability density of having $k$ nucleon-nucleon interactions, each with exactly $k_{r=1,\ldots,k}$ partonic interactions resulting with a specific (exclusive) final state $f_{r\ell}$. The last two lines in Eq.~(\ref{eq:masterdoublydiff}) simply select those final states which contain the desired particles carrying the momenta $\vec{p}^{\, a}$ and $\vec{p}^{\, b}$, and the summation over $n$ accounts for the fact that the final state can contain several a or b particles. With some combinatorics, Eq.~(\ref{eq:masterdoublydiff}) simplifies to Eq.~(\ref{eq:finaldouble}) when we identify 
\begin{align}
\frac{{\mathrm{d}}\sigma^{\rm sps}_{nn \rightarrow a + X}}{\mathrm{d}^3\vec{p}^{\, a}} & \equiv \sum_f \int \mathrm{d}{\rm PS}_f \frac{\mathrm{d}\sigma^{\rm sps}_{nn \rightarrow f}}{\mathrm{d}{\rm PS}_f} \label{eq:defsingle} \\
& \times \sum_i \delta^{\,(3)} \left(\vec p^{\, a} - \vec p^{\, a_i} \right) \times \left\{
\begin{array}{c}
 1, \ {\rm if} \ a_i \in f\\
 0, \ {\rm if} \ a_i \notin f
\end{array}
\right\} \,, \nonumber
\end{align}
and
\begin{align}
\frac{{\mathrm{d}}\sigma^{\rm sps}_{nn \rightarrow a + b + X}}{\mathrm{d}^3\vec{p}^{\, a} \mathrm{d}^3\vec{p}^{\, b}} & \equiv \sum_f \int \mathrm{d}{\rm PS}_f \frac{\mathrm{d}\sigma^{\rm sps}_{nn \rightarrow f}}{\mathrm{d}{\rm PS}_f} \\
& \times \sum_i \delta^{\,(3)} \left(\vec p^{\, a} - \vec p^{\, a_i} \right) \times \left\{
\begin{array}{c}
 1, \ {\rm if} \ a_i \in f\\
 0, \ {\rm if} \ a_i \notin f
\end{array}
\right\} \nonumber \\
& \times \sum_i \delta^{\,(3)} \left(\vec p^{\, b} - \vec p^{\, b_i} \right) \times \left\{
\begin{array}{c}
 1, \ {\rm if} \ b_i \in f\\
 0, \ {\rm if} \ b_i \notin f
\end{array}
\right\} \,. \nonumber
\end{align}
From the same formalism, also three-particle (in general $n$-particle) inclusive cross sections \cite{dEnterria:2016ids,dEnterria:2016yhy,dEnterria:2017yhd} can be derived.

\section{Perturbative-QCD framework for open heavy flavour}
\label{QCDFramework}

In this paper we will be mostly concerned in the D-meson production at $p_{\rm T} > 3\,{\rm GeV}$, which is the kinematic region considered in the LHCb double-D measurement \cite{Aaij:2012dz}. In this region the inclusive production of D mesons can be reliably described within general-mass variable-flavour-number scheme (GM-VNFS). Schematically, the cross sections are convolutions of PDFs $f_i(x,\mu_{\rm fact}^2)$, partonic cross sections $d\hat\sigma$, and fragmentation functions {(FFs)} $D_{k \rightarrow h}(z, \mu_{\rm frag}^2)$,
\begin{align}
 \mathrm{d}\sigma^{\rm sps}_{nn \rightarrow a + X} = \sum_{ijk} f_i(\mu_{\rm fact}^2) & \otimes \mathrm{d}\hat\sigma_{ij \rightarrow k + X}(\mu_{\rm fact}^2, \mu_{\rm ren}^2, \mu_{\rm frag}^2) \\
 & \otimes f_j(\mu_{\rm fact}^2) \otimes D_{k \rightarrow a}(\mu_{\rm frag}^2) \nonumber \,.
\end{align}
For single-inclusive D-meson production this has been considered at next-to-leading order (NLO) QCD first in Ref.~\cite{Kniehl:2004fy} within the so-called SACOT scheme \cite{Kramer:2000hn,Guzzi:2011ew}. In the SACOT scheme, the partonic cross sections for contributions in which the partonic subprocess is initiated by a charm quark or the fragmenting parton is a light one, are independent of the charm-quark mass $m_{\rm charm}$. This leads, in general, to diverging cross sections towards $p_{\rm T} \rightarrow 0$. In an alternative SACOT-$m_{\rm T}$ scheme \cite{Helenius:2018uul} this unphysical behaviour is resolved by accounting for the underlying kinematic constraint of heavy-quark production. In this work we use the SACOT-$m_{\rm T}$ variant, albeit in the 
considered $p_{\rm T} > 3 \, {\rm GeV}$ region, both schemes should be equivalent within the scale uncertainties. Our default choice for factorization ($\mu_{\rm fact}$), fragmentation ($\mu_{\rm frag}$) and renormalization ($\mu_{\rm ren}$)  scales is $\mu^2_{\rm fact} = \mu^2_{\rm frag} = \mu^2_{\rm ren} = {p_{\rm T}^2 + m_{\rm charm}^2}$, where $p_{\rm T}$ refers to the D-meson transverse momentum. 

The SPS contribution in which the two D mesons, $h_1$ and $h_2$, are simultaneously produced is of the form,
\begin{align}
\mathrm{d}\sigma^{\rm sps}_{\mathrm{pp} \rightarrow a + b + X} & = \sum_{ijkl} f_i(\mu_{\rm fact}^2) \otimes \mathrm{d} \hat\sigma_{ij \rightarrow k + l + X}(\mu_{\rm fact}^2, \mu_{\rm ren}^2, \mu_{\rm frag}^2) \nonumber \\
 & \otimes f_j(\mu_{\rm fact}^2) \otimes D_{k \rightarrow a}(\mu_{\rm frag}^2) \otimes D_{l\rightarrow b}(\mu_{\rm frag}^2) \,.
\end{align}
For this process, no GM-VFNS calculation is available. Thus, we will resort to the zero-mass approximation available in the NLO \textsc{diphox} \cite{Binoth:2001vm} (v.1.2) code. Taking into account the large scale uncertainties, this approximation should be sufficiently precise in the considered $p_{\rm T} > 3 \, {\rm GeV}$ region. However, the kinematical cuts applied in the considered LHCb measurement \cite{Aaij:2012dz} ($p_{\rm T} > 3 \, {\rm GeV}$ and $2 < y < 4$) include also a problematic configuration in which the two D mesons are collinear. In a full SACOT-$m_{\rm T}$ description this contribution would be finite, scaling as $\log(m_{\rm charm}^2)$ where the remaining $\log(m_{\rm charm}^2)$ terms would still need to be resummed via \emph{di-hadron FFs} \cite{Majumder:2004wh}. In a zero-mass calculation, however, the cross sections diverge in the collinear configuration. Here, as a proxy for the full SACOT-$m_{\rm T}$ treatment, we have regulated our calculations by imposing a physical cut $(\hat p_1 + \hat p_2)^2 > 4m_{\rm charm}^2$ for the fragmenting partons' four momenta $\hat p_{1,2}$. Our central choice for the QCD scales here is the average $p_{\rm T}$ of the produced two D mesons.

\begin{figure*}[htb!]
\centering
\includegraphics[width=0.495\linewidth]{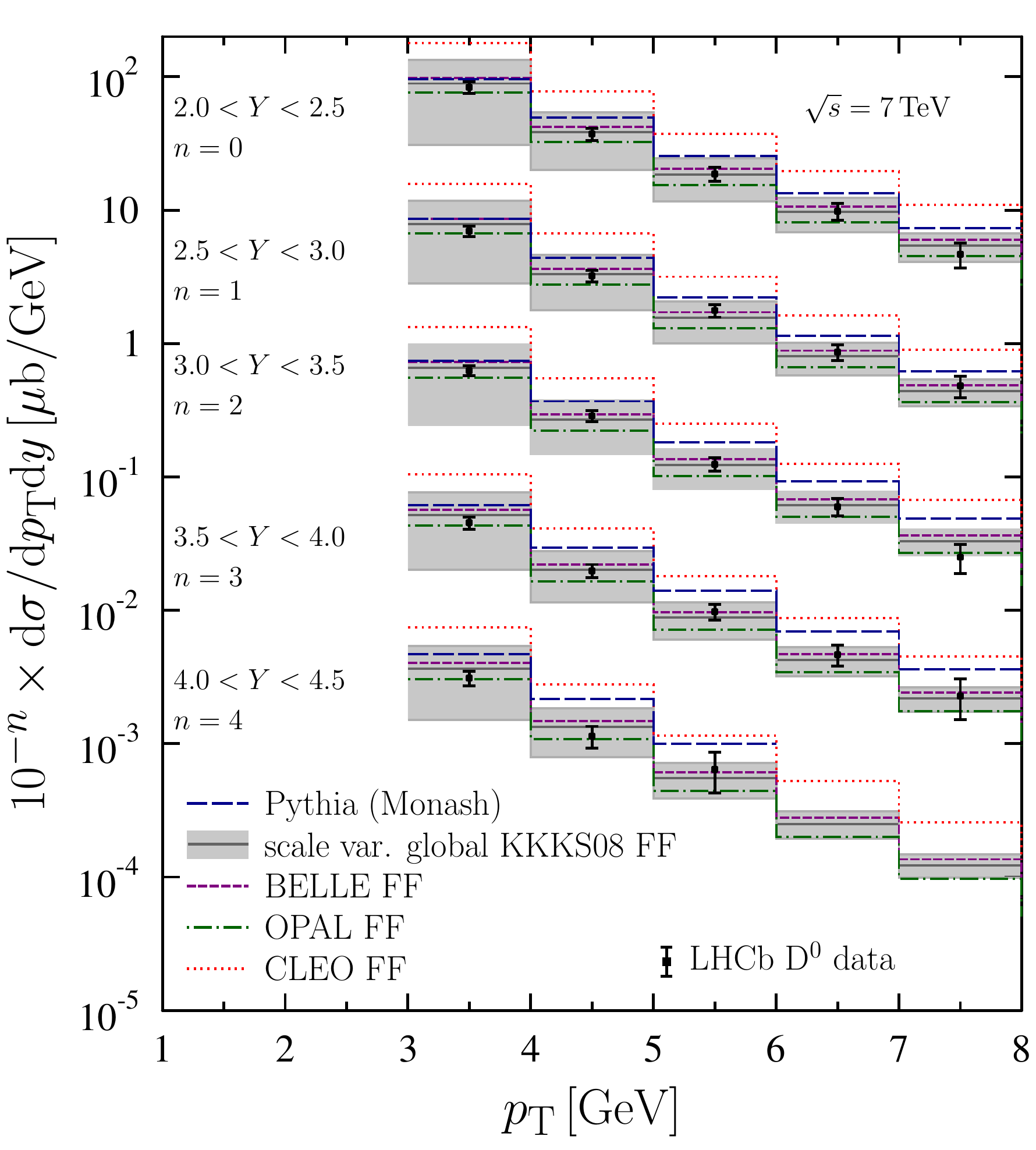}
\includegraphics[width=0.495\linewidth]{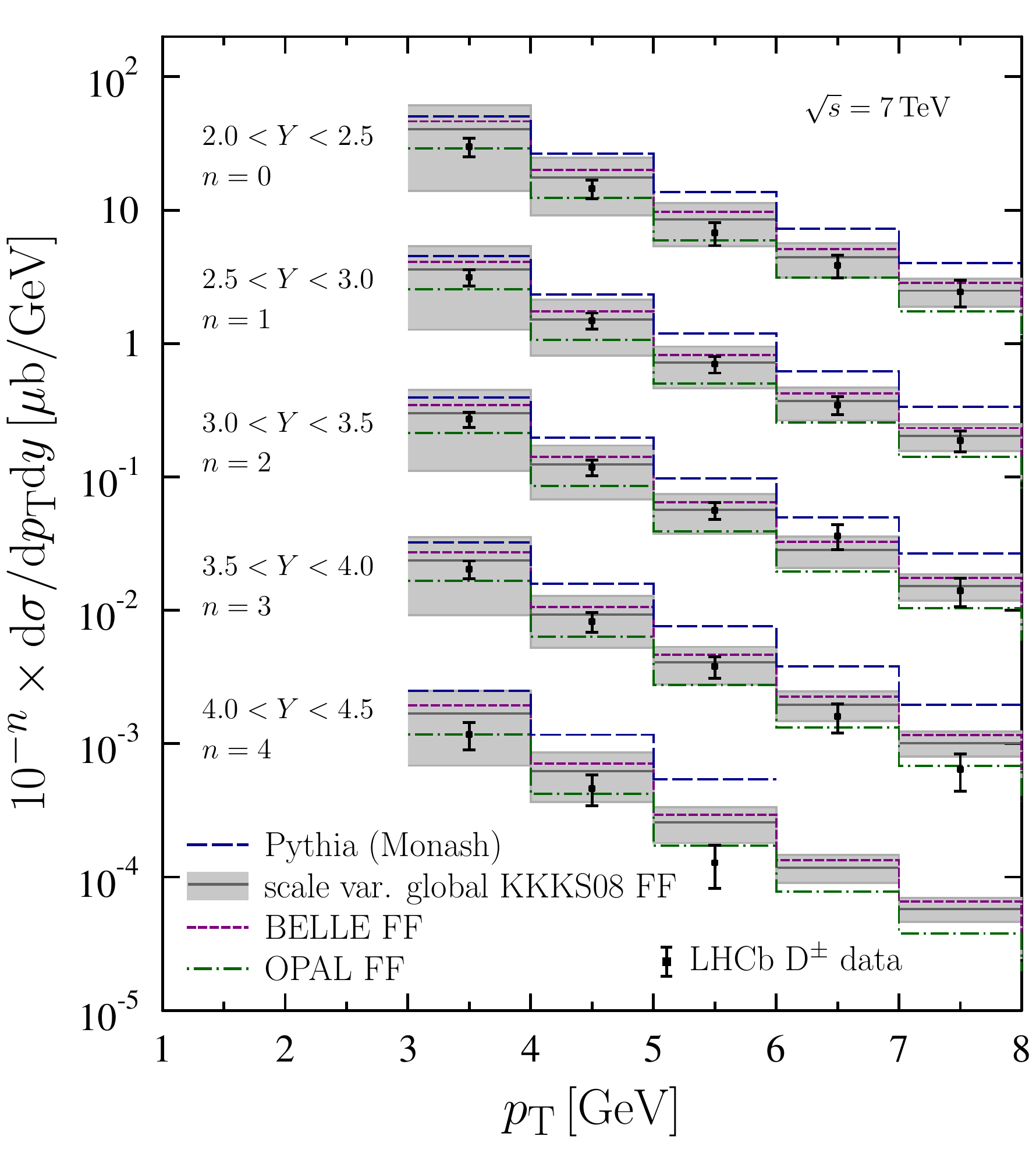}
\includegraphics[width=0.495\linewidth]{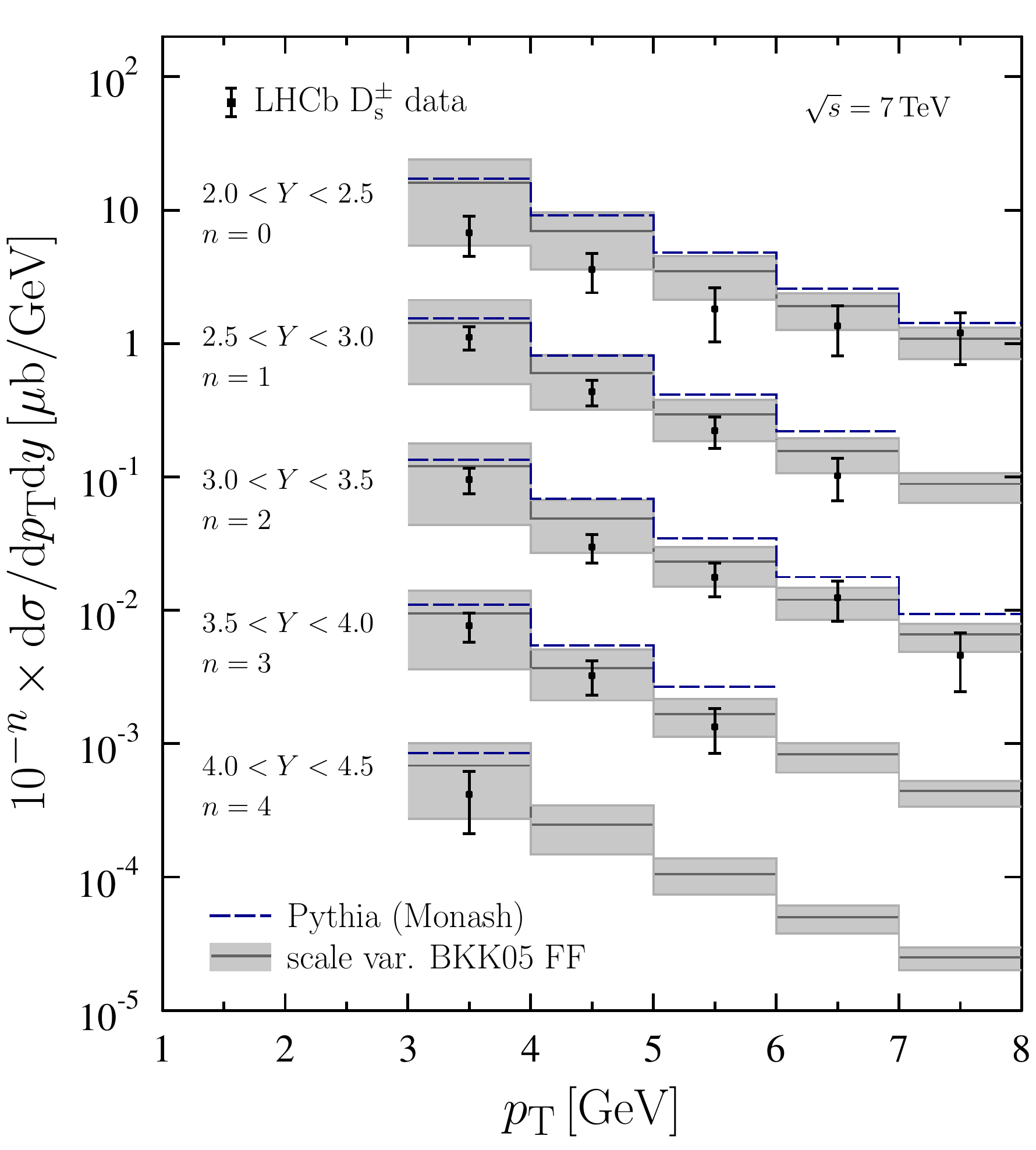}
\includegraphics[width=0.495\linewidth]{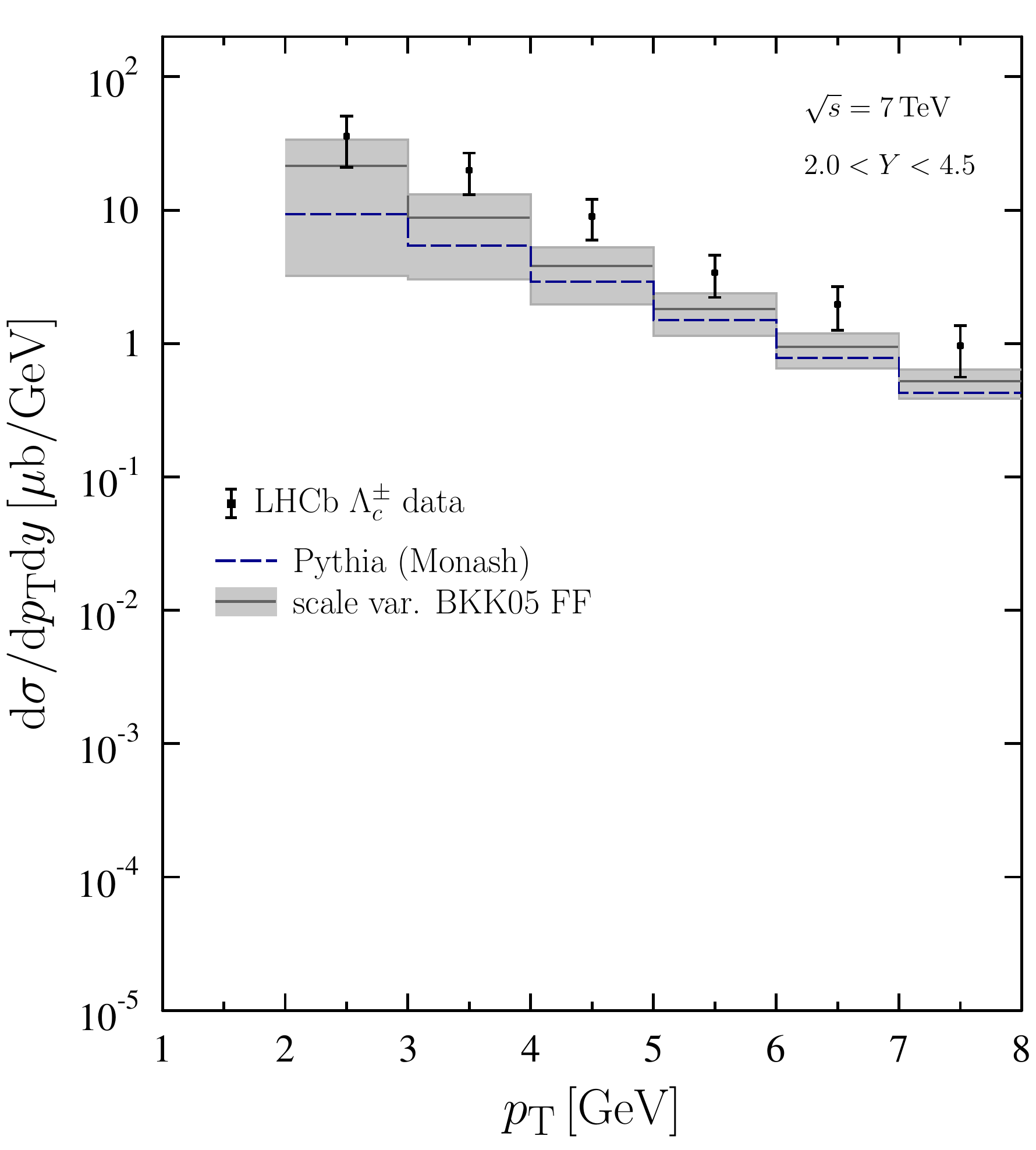}
\caption{The LHCb inclusive D-meson and $\Lambda_c^\pm$ data \cite{Aaij:2013mga} in $\sqrt{s}=7\,{\rm TeV}$ p-p collisions compared with SACOT-$m_{\rm T}$ calculation using the KKKS08 \cite{Kneesch:2007ey} and BKK05 \cite{Kniehl:2006mw} FFs. The gray bands show the scale uncertainty with central KKKS08/BKK05 FFs. The central predictions with FFs based on BELLE (purple short dashed ), OPAL (green dashed dotted), and CLEO (red dotted) data are displayed as well. The \textsc{pythia} predictions (blue long dashed) are shown also.}
\label{fig:peeteespektrit}
\end{figure*}

The dominant uncertainty in our calculations comes from the unknown higher-order (NNLO and beyond) contributions. As usual, we estimate the potential size of these corrections by varying the QCD scales as 
\begin{equation}
0.5 \leq \frac{\mu_{\rm fact}}{\mu_{\rm ren}} \leq 2, \ \ \ 0.5 \leq \frac{\mu_{\rm frag}}{\mu_{\rm ren}} \leq 2 \,,
\end{equation}
around the central scale choices and finding the combinations that give the highest and lowest prediction for each considered observable. As default, we do the scale variations in sync for the two contributions in Eq.~(\ref{eq:finaldouble}), the single-inclusive and double-inclusive SPS cross sections (17 scale configurations in total). We use NNPDF3.1pch PDFs \cite{Ball:2017nwa} in which the intrinsic charm component is zero at the mass threshold $\mu_{\rm fact} = m_{\mathrm{c}} = 1.51 \, {\rm GeV}$. The fragmentation functions for D$^0$ and D$^+$ are taken from the KKKS08 analysis \cite{Kneesch:2007ey} (see Ref.~\cite{Salajegheh:2019nea} for a very recent alternative). The KKKS08 FFs have been fitted to $e^+e^-$ data from different experiments. We have checked that while the fits to BELLE \cite{Seuster:2005tr} and OPAL \cite{Alexander:1996wy} data give essentially equally good descriptions of the inclusive LHCb D$^0$ and D$^+$ cross sections at $\sqrt{s}=7\,{\rm TeV}$ \cite{Aaij:2013mga}, the FFs fitted to CLEO data \cite{Artuso:2004pj} clearly overshoot the LHCb data at high $p_{\rm T}$. This is demonstrated in the upper panels of Figure~\ref{fig:peeteespektrit}. However, we have found that the D$^0$-to-D$^+$ ratios which are almost exclusively sensitive to the FFs are clearly best described by the OPAL variant, which also gives a better description than the BELLE FFs of the CMS midrapidity data \cite{Sirunyan:2017xss} at very-high $p_{\rm T}$ \cite{HPtalk}. Thus, in this paper, we adopt the OPAL FFs from the KKKS08 package. For D$_{\mathrm{s}}^\pm$ and $\Lambda_{\mathrm{c}}^\pm$ FFs we use BKK05 \cite{Kniehl:2006mw} analysis. While the LHCb and ALICE single-inclusive D$_{\rm s}^\pm$ data \cite{Aaij:2013mga,Acharya:2019mgn} are well consistent with these FFs, the $\Lambda_{\mathrm{c}}^\pm$ data \cite{Aaij:2013mga} are underestimated by the BKK05 $\Lambda_{\mathrm{c}}^\pm$ FFs. Our comparisons with the LHCb data on $D_{\mathrm{s}}^\pm$ and $\Lambda_{\mathrm{c}}^\pm$ are shown in the bottom panels of Figure~\ref{fig:peeteespektrit}.

The KKKS08 and BKK05 FFs do not discriminate between charge-conjugate states, but are given as a sum. 
(e.g. $D^{{\rm D}^0 + \overline{\rm D^0}}_{i}$). 
In what follows, however, we will need the D-meson FFs one by one. Taking the D$^0$ states here as an example, we will use the following prescription for the charm-quark containing state,
\begin{align}
D^{\rm D^0}_{\mathrm{c}} & = D^{{\rm D}^0 + \overline{\rm D^0}}_{\mathrm{c}/\overline{\mathrm{c}}} \,, \\
D^{\rm D^0}_{\overline{\mathrm{c}}} & = 0 \,, \\
D^{\rm D^0}_i & = \frac{1}{2} D^{{\rm D}^0 + \overline{\rm D^0}}_{i}, \ {i \neq \mathrm{c}/\overline{\mathrm{c}}} \,,
\end{align}
and an analogous one for the antiquark-containing state,
\begin{align}
D^{\overline{\rm D}^0}_{\mathrm{c}} & = 0 \,, \\
D^{\overline{\rm D}^0}_{\overline{\mathrm{c}}} & = D^{{\rm D}^0 + \overline{\rm D^0}}_{\mathrm{c}/\overline{\mathrm{c}}}  \,, \\
D^{\overline{\rm D}^0}_i & = \frac{1}{2} D^{{\rm D}^0 + \overline{\rm D^0}}_{i}, \ {i \neq \mathrm{c}/\overline{\mathrm{c}}} \,.
\end{align}

In addition to the NLO QCD framework described above, we present the predictions from \textsc{Pythia} 8 Monte-Carlo event generator using the standard ``Monash 2013 tune'' \cite{Skands:2014pea}. A sample of minimum-bias events were generated, including also MPIs, from which the different D-meson combinations within the LHCb acceptance were picked up to obtain the cross sections for each pair. In line with the LHCb measurements, each pair of D mesons is counted separately. In Figure~\ref{fig:peeteespektrit} we also show the \textsc{Pythia} predictions for the inclusive D mesons and $\Lambda_{\mathrm{c}}^\pm$, generated with the provided Rivet analysis \cite{Buckley:2010ar}. In general, the \textsc{Pythia} setup overpredicts the LHCb D-meson measurements, and the disagreement is stronger for D$^\pm$ and D$^\pm_{\mathrm{s}}$ than for D$^0$. A similar behaviour has been recently observed in the case of jets containing a ${\rm D}^0$ meson \cite{Acharya:2019zup}. The measured $\Lambda_{\mathrm{c}}^\pm$ cross sections are, in turn, underestimated by \textsc{Pythia}. In the Monash tune the parameters related to charm fragmentation were constrained using a limited set of LEP data. Partly the interpretation of these data is hindered by the large feed-down from B-mesons. Furthermore, the data is not sensitive to $\mathrm{g} \rightarrow \mathrm{c\overline{c}}$ branchings that are abundant at the LHC. Thus the observed disagreement could potentially be cured by re-tuning the relevant parameters using a larger sample of charm-production data from LEP, HERA and LHC.

\section{Results}
\label{Results}

We will now compare our results for double D-meson production with the LHCb p-p data \cite{Aaij:2012dz}, and make predictions for p-Pb collisions. As for $\sigma_{\rm eff}$, we will consider the variation $10~\text{mb} < \sigma_{\rm eff} < 25~\text{mb}$ which is roughly the range deduced from jet, W$^\pm$ and photon measurements \cite{Aaboud:2018tiq}. The uncertainty estimates shown in the plots combine the scale uncertainty and the variation in $\sigma_{\rm eff}$.

\begin{figure}[htb!]
\centering
\includegraphics[width=0.95\linewidth]{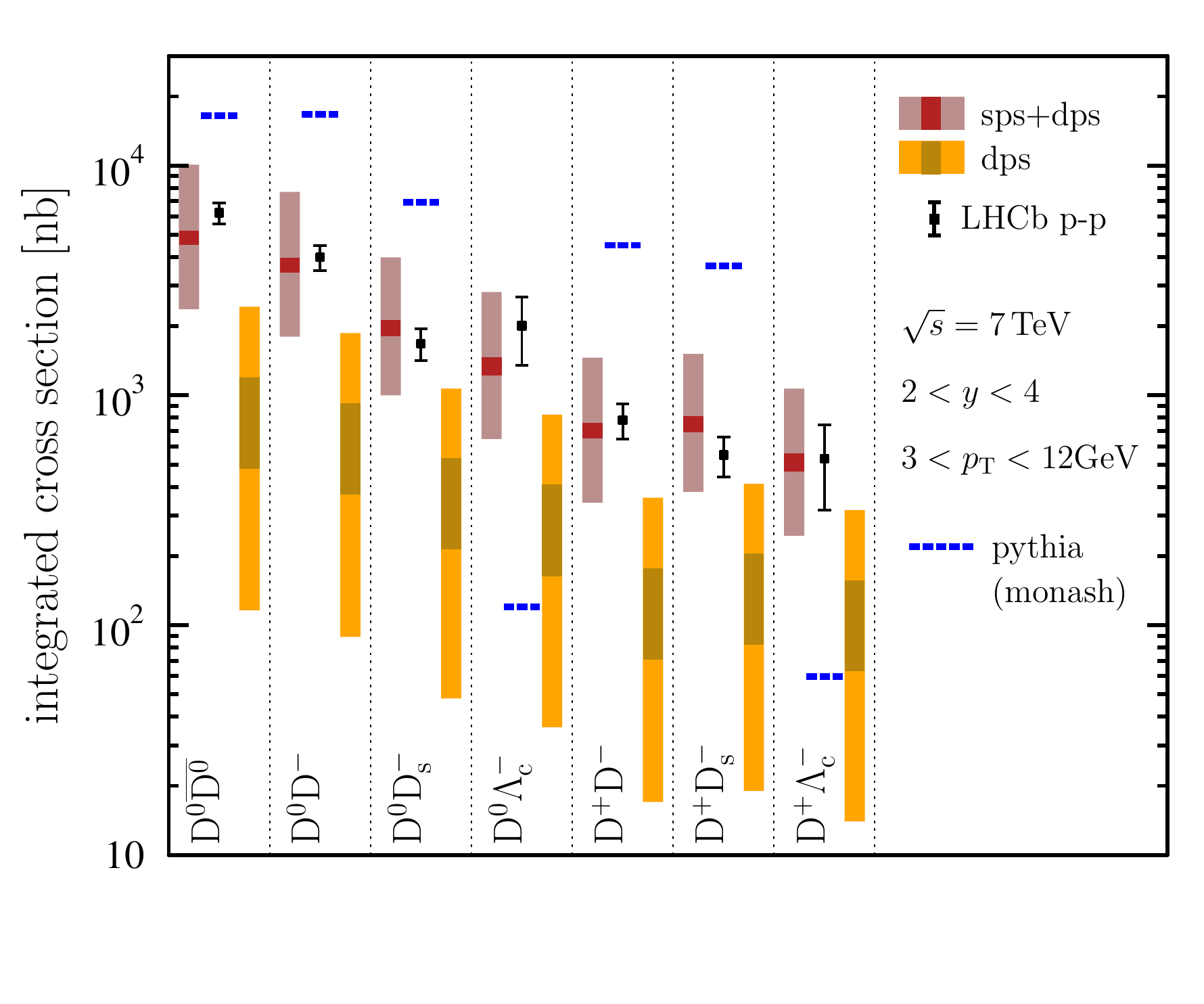}

\vspace{-1.1cm}
\includegraphics[width=0.95\linewidth]{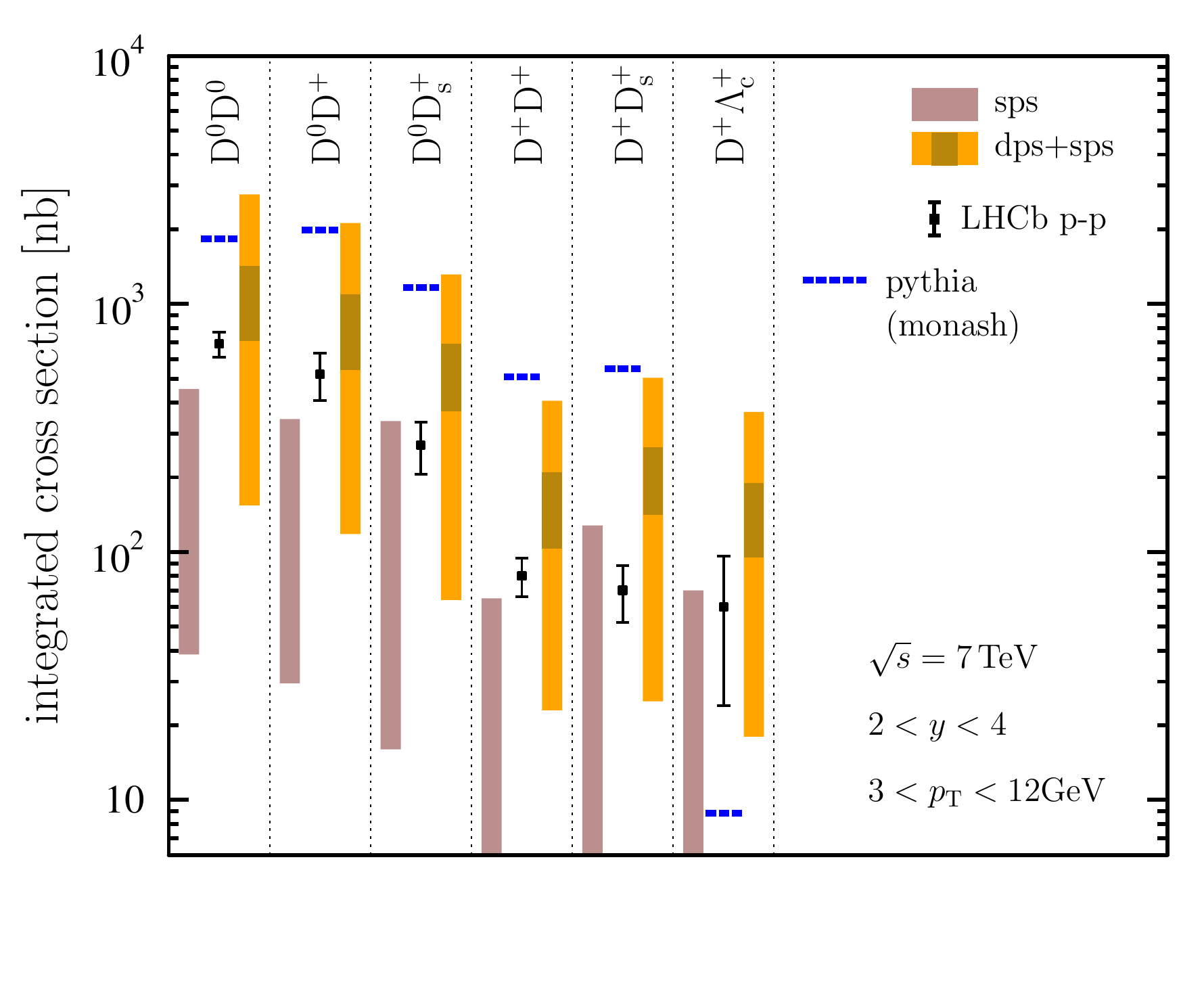}
\caption{The integrated double-D cross sections for opposite-sign (upper panel) and like-sign (lower panel) cases. The coloured bands denote the combined scale and $\sigma_{\rm eff}$ uncertainty in NLO calculations. The inner darker bands include only the variation in $\sigma_{\rm eff}$. The \textsc{Pythia} predictions are shown as blue dashed lines. The data are from Ref.~\cite{Aaij:2012dz}.}
\label{fig:ppyields}
\end{figure}

\subsection{p-p collisions}
\label{ppResults}

In the case of p-p collision, Eqs.~(\ref{eq:finaldouble}) and (\ref{eq:sigmaeffAB}) reduce to
\begin{equation}
 \frac{{\mathrm{d}}\sigma_{\mathrm{pp} \rightarrow a + b + X}}{\mathrm{d}\vec{p}^{\, a} \mathrm{d}\vec{p}^{\, b}} = \label{eq:doublepp}
 \frac{{\mathrm{d}}\sigma^{\rm sps}_{\mathrm{pp} \rightarrow a + b + X}}{\mathrm{d}\vec{p}^{\, a} \mathrm{d}\vec{p}^{\, b}} \ + 
\frac{m}{\sigma_{\rm eff}} \frac{{\mathrm{d}}\sigma^{\rm sps}_{\mathrm{pp} \rightarrow a + X}}{\mathrm{d} \vec{p}^{\,a} } \frac{{\mathrm{d}}\sigma^{\rm sps}_{\mathrm{pp} \rightarrow b + X}}{\mathrm{d} \vec{p}^{\,b} } \,.
\end{equation}
Our results for the integrated cross sections within the LHCb acceptance are shown in Figure~\ref{fig:ppyields}. For the opposite-sign D mesons (upper panel), the SPS contribution is clearly larger than the DPS one, and the agreement with the data is very good. The measured systematics among different combinations of D (and $\Lambda_{\rm c}$) species is well reproduced by the used set of FFs. As the cross section accumulates from the lower end of the considered $p_{\mathrm{T}}$ range, the scale uncertainty is sizable and dominates over the variation in $\sigma_{\rm eff}$. For like-sign final states (lower panel) the DPS becomes the dominant production mechanism. Again, the calculation agrees with the data within the scale uncertainties, though our central scale choice seem to somewhat overestimate the cross sections. The disagreement between the data and \textsc{Pythia} results is considerably larger than in the single-inclusive case (Figure~\ref{fig:peeteespektrit}). Apart from pairs including $\Lambda_{\rm c}$ the predicted cross sections are 4--8 times higher than the data. For pairs including $\Lambda_{\rm c}$ the systematic is again the opposite.

\begin{figure}[htb!]
\centering
\includegraphics[width=0.99\linewidth]{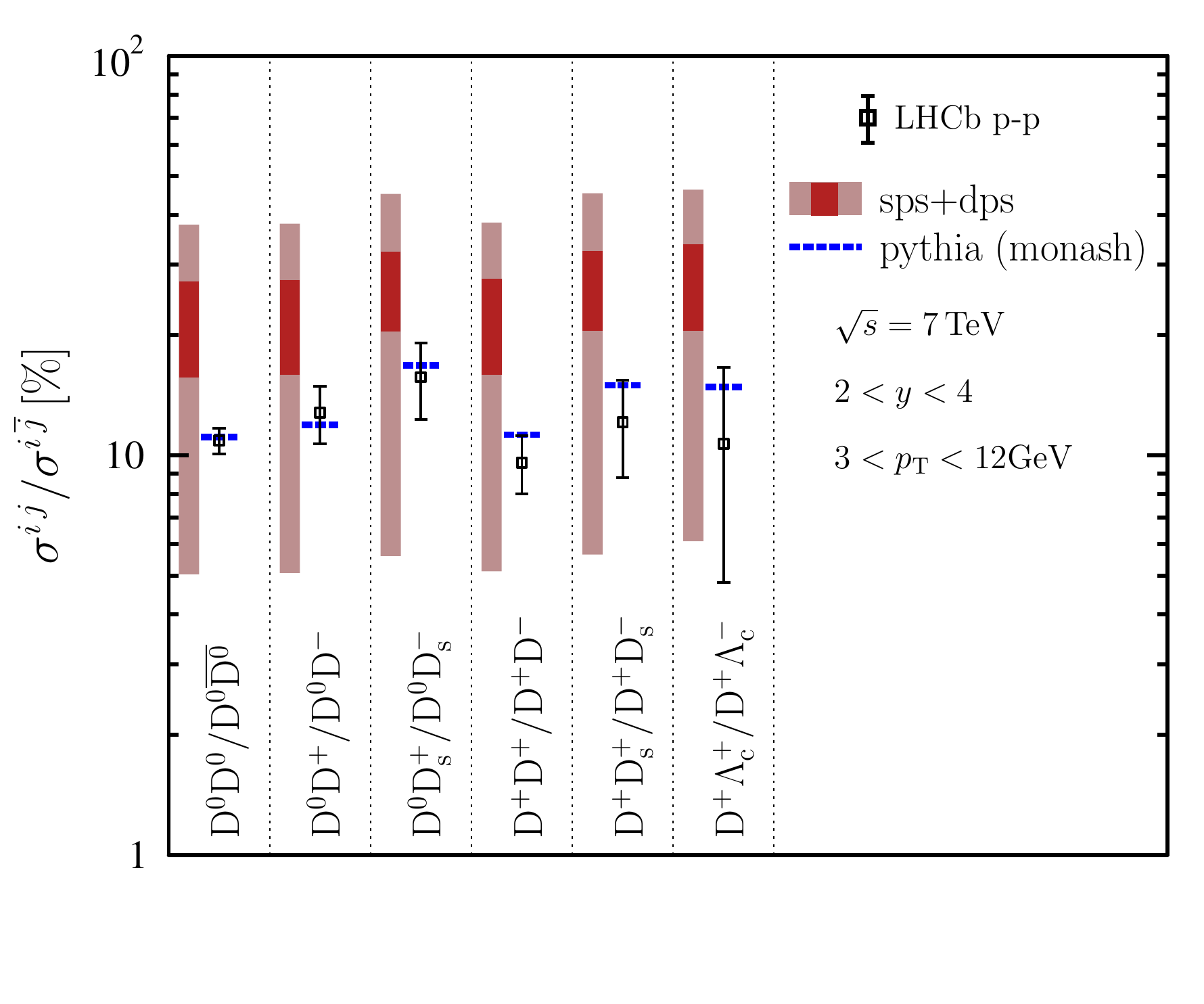}

\vspace{-1.2cm}
\includegraphics[width=0.99\linewidth]{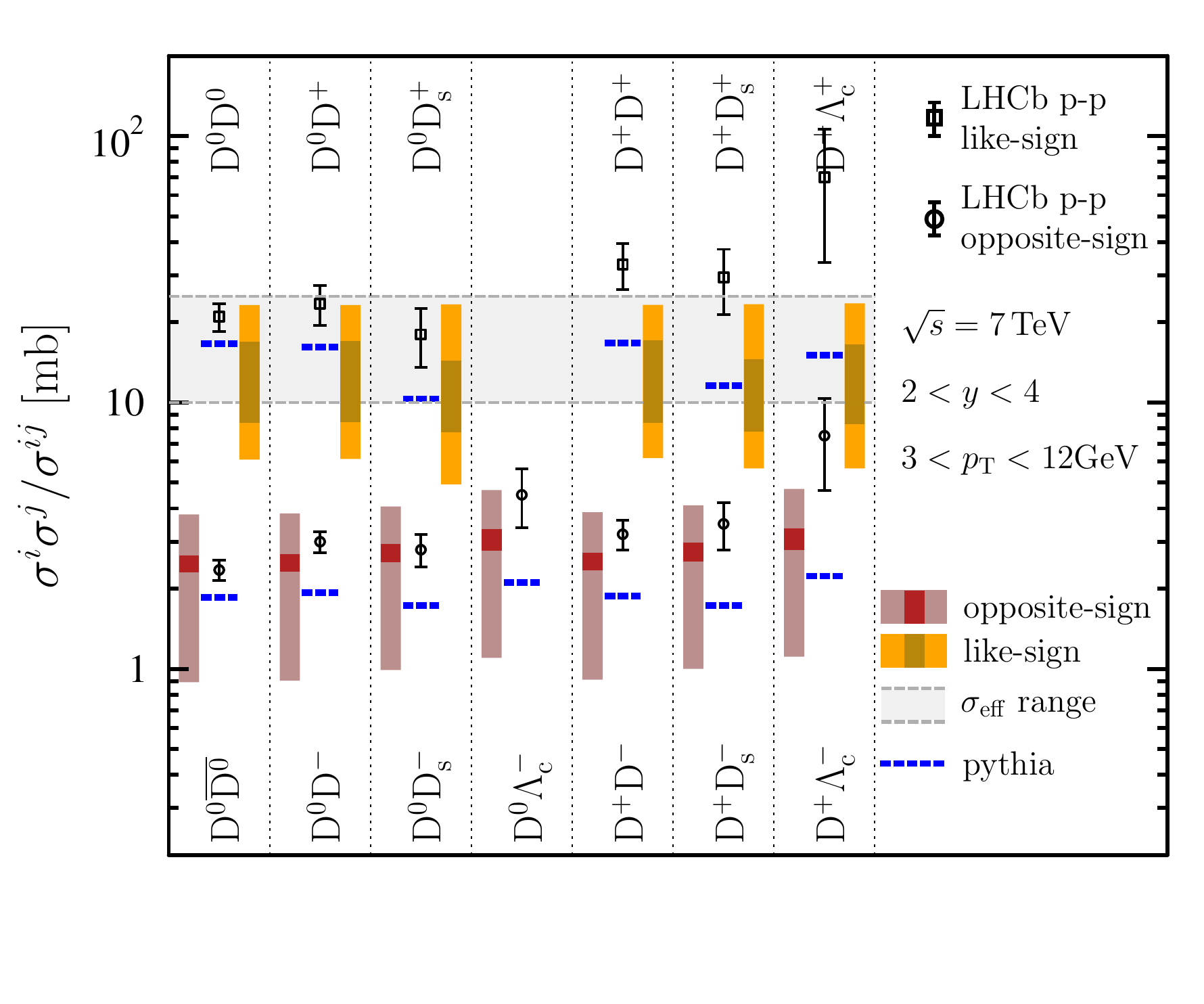}

\caption{Upper panel: like-sign vs. opposite-sign ratios, see Eq.~(\ref{eq:kivasuhde2}). Lower panel: Product of two single-inclusive D-meson cross sections divided by the double-D cross sections, see Eq.~(\ref{eq:kivasuhde}). The coloured bars denote the combined scale and $\sigma_{\rm eff}$ uncertainty, and the inner darker bands include only the variation in $\sigma_{\rm eff}$. The dashed lines correspond to what \textsc{Pythia} predicts. The upper set of bands/lines/data points correspond to like-sign D mesons, and the lower set to opposite-sign combinations.
}
\label{fig:ppratios}
\end{figure}

More insight can be obtained from Figure~\ref{fig:ppratios} where we show cross-section ratios. The upper panel shows ratios between the double like-sign vs. opposite-sign cross sections,
\begin{equation}
{\sigma^{ab}}/{\sigma^{a\overline{b}}} \equiv \frac{\sigma_{\mathrm{pp} \rightarrow a + b + X}}{\sigma_{\mathrm{pp} \rightarrow a + \overline{b} + X}} \,. \label{eq:kivasuhde2}
\end{equation}
These measure essentially the ratio between the DPS and SPS contributions. There is clearly a fair data-to-theory agreement within the scale and $\sigma_{\rm eff}$ uncertainties. Our central predictions somewhat overestimate the measured values which is consistent with Figure~\ref{fig:ppyields}. The scale uncertainties do not cancel out since the partonic channels for like-sign and opposite-sign production are different (e.g. $\rm c \overline{\rm c}$ pair production is significant for $\rm D^0\overline{D^0}$ final state but not for $\rm D^0D^0$). Interestingly, the \textsc{Pythia} results are in excellent agreement with the LHCb data even though the absolute cross sections are way off. 
Since the numerator in Eq.~(\ref{eq:kivasuhde2}) is sensitive to DPS (or MPIs in general), we conclude that the good agreement here suggest that the inconsistencies observed in Figure~\ref{fig:ppyields} are indeed due to poorly-constrained charm fragmentation, rather than the MPI modelling in \textsc{Pythia} \cite{Sjostrand:1987su,Sjostrand:2004pf,Sjostrand:2017cdm}.

The bottom panel of Figure~\ref{fig:ppratios} shows ratios
\begin{align}
{\sigma^a\sigma^b}/{\sigma^{ab}} & \equiv 
m \,  \frac{\sigma_{\mathrm{pp} \rightarrow a + X} \times \sigma_{\mathrm{pp} \rightarrow b + X}}{\sigma_{\mathrm{pp} \rightarrow a + b + X}}
\,. \label{eq:kivasuhde}
\end{align}
From Eq.~(\ref{eq:doublepp}) we see that in the absence of SPS, this ratio would be equal to $\sigma_{\rm eff}$, but if there is a contribution from SPS, the ratio will be below $\sigma_{\rm eff}$. 
In general, our predictions for the opposite-sign case match very well with the data, but tend to underestimate the measured like-sign ratios. 
This is well in line with our earlier observations and also here a better overall agreement would be obtained if the DPS cross section would be somewhat smaller. Thus, the double-charm production data would prefer a somewhat larger phenomenological $\sigma_{\rm eff}$ than what other measurements indicate \cite{Aaboud:2018tiq}.  The \textsc{Pythia} predictions are here well compatible with our NLO calculations, though they somewhat undershoot the measured ratios both for like- and opposite-sign ratios. This further supports our conclusion that the disagreement observed in Figures~\ref{fig:peeteespektrit} and \ref{fig:ppyields} arise from the fragmentation scheme in \textsc{Pythia}.

\subsection{p-Pb collisions}
\label{pPbResults}

\begin{figure}[htb!]
\centering
\includegraphics[width=0.99\linewidth]{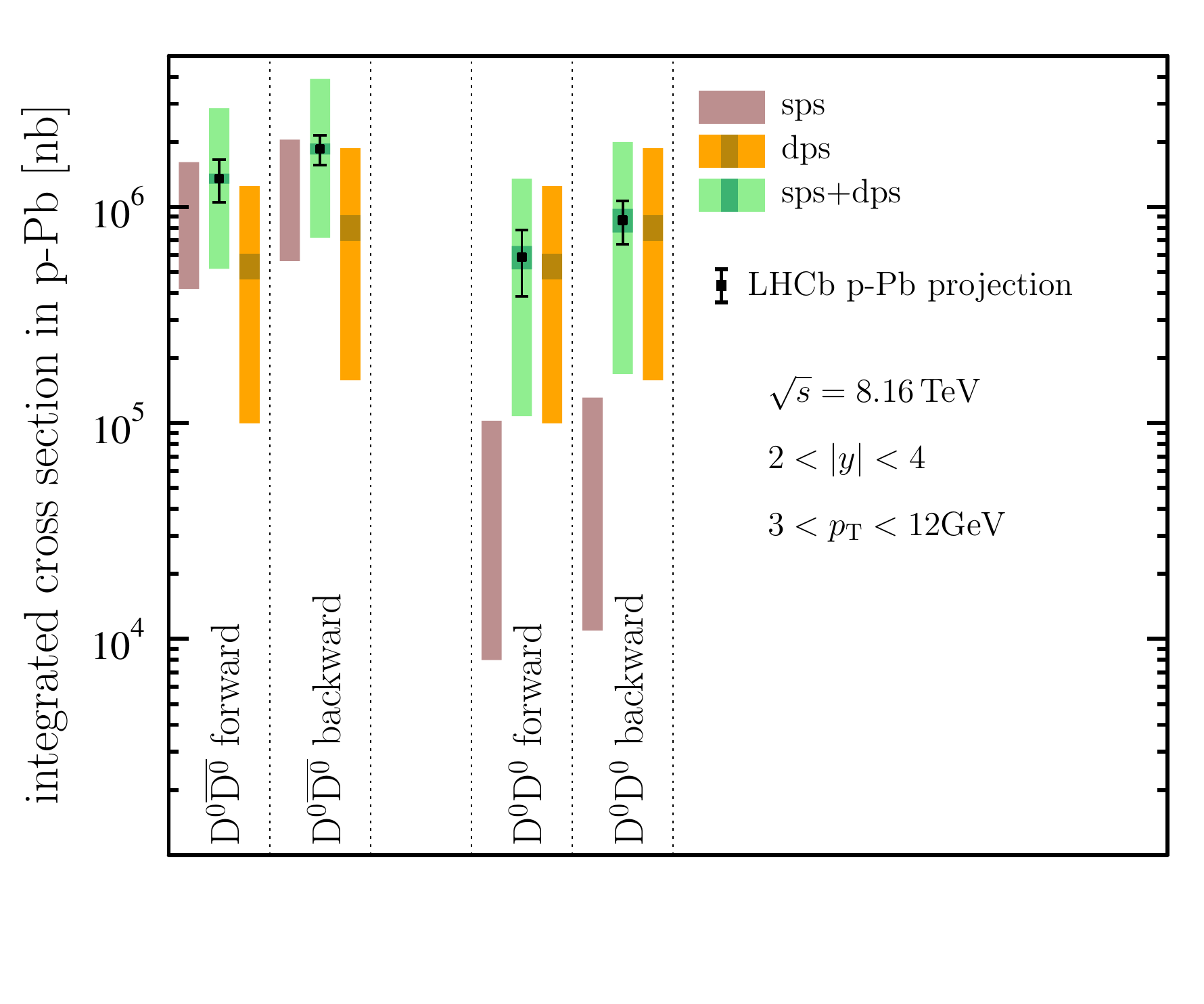}

\vspace{-1.2cm}
\includegraphics[width=0.99\linewidth]{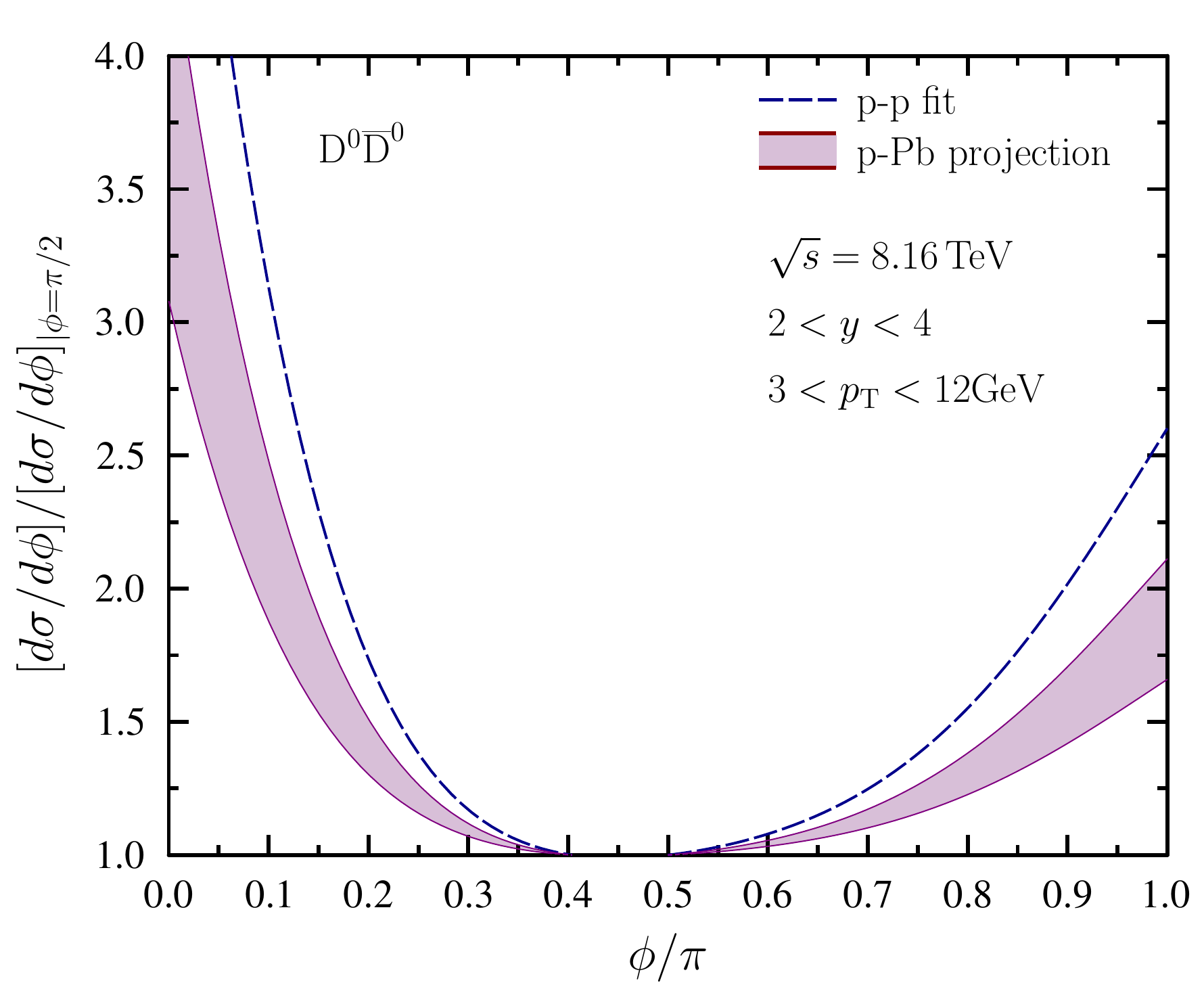}
\caption{Upper panel: Integrated cross sections for ${\rm D}^0\overline{\rm D^0}$ and ${\rm D}^0{\rm D}^0$ cross sections in p-Pb collisions within the LHCb acceptance at $\sqrt{s}=8.16 \, {\rm TeV}$. The coloured bars denote the combined scale and $\sigma_{\rm eff}$ uncertainty, and the inner darker bands include only the variation in $\sigma_{\rm eff}$. The LHCb projections correspond to $12.2\,{\rm nb}^{-1}$ (forward) and $18.6\,{\rm nb}^{-1}$ (backward) luminosities assuming the predicted central value and overall efficiency of $1.2 \times 10^{-6}$. Lower panel: A sketch of the relative azimuthal-angle dependence in p-p and projected p-Pb collisions.}
\label{fig:pPbyields}
\end{figure}

The reasonble description of the p-p data gives us confidence to apply the framework in p-$A$ collisions. In this case, Eqs.~(\ref{eq:finaldouble}) and (\ref{eq:sigmaeffAB}) reduce to
\begin{align}
& \frac{{\mathrm{d}}\sigma_{\mathrm{p}A \rightarrow a + b + X}}{\mathrm{d}\vec{p}^{\, a} \mathrm{d}\vec{p}^{\, b}} = \label{eq:doublepA} 
A  \left[ \frac{{\mathrm{d}}\sigma^{\rm sps}_{nn \rightarrow a + b + X}}{\mathrm{d}\vec{p}^{\, a} \mathrm{d}\vec{p}^{\, b}} \ +  
\frac{m}{\sigma_{\rm eff}^{\mathrm{p}A}} \frac{{\mathrm{d}}\sigma^{\rm sps}_{nn \rightarrow a + X}}{\mathrm{d} \vec{p}^{\,a} } \frac{{\mathrm{d}}\sigma^{\rm sps}_{nn \rightarrow b + X}}{\mathrm{d} \vec{p}^{\,b} } \right] 
\end{align}
with
\begin{align}
\frac{1}{\sigma_{\rm eff}^{\mathrm{p}A}} \equiv \frac{ 1}{\sigma_{\rm eff}} \times  \Bigg\{1  
&  + \sigma_{\rm eff} \frac{A-1}{A^2} \int \mathrm{d}^2\vec{B} \left[ T_{\mathrm{n}A}( \vec{B}\,) \right]^2  \Bigg\} \label{eq:sigmaeffpA} 
\,.
\end{align}
The impact-parameter integral for $A=208$ (Pb) gives 
\begin{equation}
\int \mathrm{d}^2\vec{B} \left[ T_{\rm nPb}\left( \vec{B}\right) \right]^2 \approx 31.66~{\rm mb}^{-1}
\end{equation}
taking $d=0.54\,{\rm fm}$ and $r=6.49\,{\rm fm}$ in the Woods-Saxon profile,
\begin{equation}
\rho^A(\vec s, z) = {n_0}\left[{1 + \exp \left(\frac{\sqrt{\vec s^{\,2}+z^2} - r}{d}\right)}\right]^{-1} \,,
\end{equation}
and fixing $n_0$ by the normalization condition of Eq.~(\ref{eq:norm}). With $\sigma_{\rm eff} = 10\ldots 25 ~{\rm mb}$, we find
\begin{align}
\frac{1}{\sigma_{\rm eff}^{\mathrm{p{Pb}}}} \approx \frac{2.5 \ldots 4.8}{\sigma_{\rm eff}} 
\end{align}
in full consistency e.g. with Ref.~\cite{dEnterria:2012jam}. That is, the DPS signal is enhanced approximately by a factor of three in comparison to p-p scattering. Our results for the integrated cross sections within the LHCb kinematics are shown in Figure~\ref{fig:pPbyields}. Here, we have only considered D$^0$ production which has the largest cross sections, see Figure~\ref{fig:ppyields}, and the $y$ acceptance refers to that in the center-of-mass frame of the p-Pb collision. When computing the per-nucleon cross sections $\sigma^{\rm sps}_{nn \rightarrow a + b + X}$ and $\sigma^{\rm sps}_{nn \rightarrow a/b + X}$, we have used the EPPS16 nuclear modifications \cite{Eskola:2016oht} for Pb. At the LHCb kinematics this leads to a $\sim 20\%$ suppression for p-Pb (forward) SPS cross sections, but since this is squared in DPS contribution, the suppression can reach $\sim 40\%$ in DPS case. For Pb-p configuration (backward) the nuclear-PDF effects are smaller. In comparison to the p-p case in Figure~\ref{fig:ppyields} the impact of enhanced DPS contribution is clear: Whereas in p-p case the DPS contribution to the opposite-sign yield was rather small in comparsion to SPS, in p-Pb collisions the two are comparable. For the like-sign yields the SPS contribution in p-Pb collisions is entirely overpowered by the DPS part, whereas in the p-p case the SPS still had a 20\% contribution or so. Due to the additional contribution from the $T^2_{{\rm n}A}(\vec B)$ integral in Eq.~(\ref{eq:sigmaeffpA}), the variation in $\sigma_{\rm eff}$ plays only a minor role as indicated in Figure~\ref{fig:pPbyields}. The $\sim$30\% differences between forward and backward cross sections are due to the EPPS16 nuclear effects. Thus, by a suitable measurement where other theoretical uncertainties would cancel out, e.g. a forward-to-backward ratio for double D-meson production, further constraints for nuclear PDFs could, perhaps, be obtained.

An interesting question is whether these cross sections are large enough to be measured with the already collected Run-II data. In Run-II data taking the luminosities collected by the LHCb were $12.2\,{\rm nb}^{-1}$ for p-Pb (forward) and $18.6\,{\rm nb}^{-1}$ Pb-p (backward) collisions \cite{Aaij:2019lkm}. The overall detection efficiency $\epsilon$ for ${\rm D}^0\overline{{\rm D^0}}$ and ${\rm D}^0{{\rm D}}^0$ final states in the LHCb p-p measurement \cite{Aaij:2012dz} was approximately $\epsilon \approx 1.2 \times 10^{-6}$. Using these luminosities and efficiencies with our central theoretical predictions we calculate the expected number of events $N$ from which the statistical uncertainty is obtained as $\sqrt{N}/N$. These estimates are also shown in Figure~\ref{fig:pPbyields}. Within the scale uncertainties we expect approximately $10\dots40$ ${\rm D}^0\overline{{\rm D}}^0$ pairs in p-Pb collisions (forward), and $20\ldots80$ in Pb-p configuration (backward). For the like-sign case the corresponding numbers are $2\dots20$ ${\rm D}^0{\rm D}^0$ pairs in p-Pb collisions (forward), and $4\ldots40$ in Pb-p configuration (backward). Thus, we are led to conclude that the double D-meson production -- at least the opposite-sign case -- should be observable at the LHCb with the Run-II luminosity. Lowering the minimum-$p_{\rm T}$ cut below $3\,{\rm GeV}$ would easily increase the yields to a definitely measurable level, but towards lower $p_{\rm T}$ our predictions become increasingly uncertain.

As is well known, the increased importance of the DPS contribution in p-Pb collisions may significantly affect the kinematic distributions \cite{Strikman:2010bg,Lappi:2012nh}. Particularly interesting observable is the relative azimuthal-angle $\Delta\phi$ distribution of the two D mesons \cite{Vogt:2018oje}. In p-p collisions \cite{Aaij:2012dz} the $\Delta\phi$ distribution for ${\rm D}^0\overline{{\rm D^0}}$ peaks at $\Delta\phi=0$ for the logarithmically enhanced $g \rightarrow c\overline{c}$ splitting, and at $\Delta\phi=\pi$ due to the leading-order contributions that are back-to-back in transverse plane. These are commonly referred to as the \emph{near-side peak} and the \emph{away-side peak}, respectively. The disappearance of the away-side peak has long been predicted to be the smoking gun of saturation physics \cite{Albacete:2010pg,Albacete:2014fwa}. However, the enhanced DPS contribution in p-Pb collisions will generate a $\Delta\phi$-independent contribution which levels off these peaks. Unfortunately our NLO QCD framework cannot reliably predict the $\Delta\phi$ dependence near $\Delta\phi=\pi$ 
but a soft-gluon resummation encoded e.g. in parton showers, would be required. To estimate the effect, we have fitted the $\Delta\phi$ dependence of the LHCb ${\rm D}^0\overline{{\rm D}^0}$ data \cite{Aaij:2012dz} in p-p collisions assuming a negligible contribution from DPS. This assumption is consistent both with our results (see Figure~\ref{fig:ppyields}) and also with Ref.~\cite{Vogt:2018oje} where it has been shown that for $0 \lesssim \Delta\phi \lesssim \pi/2$ the fixed-order QCD quite correctly predicts the $\Delta\phi$ dependence. Our sketchy estimate for p-Pb is then a linear combination 
\begin{equation}
\frac{\mathrm{d}\sigma^{\rm pPb}}{\mathrm{d}\Delta\phi} \propto \frac{\mathrm{d}\sigma^{\rm pp}}{\mathrm{d}\Delta\phi}{\big|}_{\rm fitted} + \beta
\end{equation}
where the constant $\beta$ is determined by the relative importance of integrated SPS and DPS cross sections. Depending on the scale choices, we estimate the DPS contribution to be roughly between 20\%\ldots40\%. The resulting projection for the $\Delta\phi$ dependence is shown in the lower panel of Figure~\ref{fig:pPbyields} where the coloured band comes from the scale and $\sigma_{\rm eff}$ uncertainties. We see that both the near- and away-side peaks become less pronounced in p-Pb than what they are in p-p. Thus, we can confirm that the DPS contributions should be considered when interpreting the possible (probable?) weakening of the away-side two-particle correlations in terms of e.g. saturation physics. Since the DPS contribution should be nearly the same for ${\rm D}^0\overline{{\rm D}^0}$ and ${\rm D}^0{\rm D^0}$ final states (in our calculations they are equal), the difference
\begin{equation}
\frac{\mathrm{d}\sigma_{\mathrm{pPb} \rightarrow {\rm D}^0\overline{{\rm D}^0}+X}}{\mathrm{d}\Delta\phi} - \frac{\mathrm{d}\sigma_{\mathrm{pPb} \rightarrow {\rm D}^0{\rm D}^0 + X}}{\mathrm{d}\Delta\phi} \approx \frac{\mathrm{d}\sigma^{\mathrm{sps}}_{\mathrm{pPb} \rightarrow {\rm D}^0\overline{{\rm D}^0}+ X}}{\mathrm{d}\Delta\phi}
\end{equation}
should serve to subtract the ``pedestal'' DPS yield in a rather model-independent way and, as indicated above, correspond very closely to the SPS contribution in ${\rm D}^0\overline{{\rm D}^0}$ production. Even though the presented calculations are for double D-meson production, we would expect a similar reduction of the near- and away-side peaks also for light-flavour hadrons (such as $\pi^+ \pi^-$) in due to enhanced DPS contribution in p-Pb collisions.

\section{Summary}

We have explored the double-inclusive D-meson production at the LHC with focus on the forward LHCb kinematics. The contributions of double-parton scatterings were included in the approximation of independent parton-parton collisions, and the required single-parton cross sections were computed within the collinearly factorized QCD at an NLO level. We confronted our predictions with the LHCb p-p data finding a good, or least an acceptable agreement within the QCD scale uncertainties and reasonable variation in the effective cross $\sigma_{\rm eff}$. As a whole, the LHCb data would prefer a rather large $\sigma_{\rm eff}$ compared to values derived from other final states. We also compared the LHCb p-p data with \textsc{Pythia} predictions. We found that the absolute cross sections for single- and double-inclusive open-charm production are not well reproduced by the widely used Monash tune. However, the cross-section ratios, which are less sensitive to the details in charm fragmentation, are described equally well or even better than what our NLO calculations do. Since the ratios are more sensitive to the multi-parton dynamics than the heavy-quark fragmentation model, it seems that the latter will need some further tuning to establish an agreement with the absolute cross sections.

In addition, we applied our framework to the case of p-Pb collisions in which the contribution from double-parton scattering is predicted to get significantly enhanced due to multiple nucleon-nucleon interactions. Our calculations, accounting for realistic reconstruction efficiencies, indicate that the yields should be high enough to be measured with the already-collected LHC Run-II data at the LHCb. This should provide a clear evidence for the hard double-parton scattering in p-Pb. As the contributions from single- and double-parton scatterings to opposite-sign double-D pair become comparable, also the azimuthal correlations are significantly altered. Therefore it seems necessary to take the double-parton scattering component into account when interpreting e.g. the possible complete or probable partial disappearance of the away-side peak.

\section*{Acknowledgments}

We thank Michael Winn for clarifying us details of the LHCb p-p measurements and Peter Skands for discussions related to the Monash-tune applied in \textsc{Pythia} simulations. Our work was financed by the Academy of Finland, project n.o.~308301. The Finnish IT Center for Science (CSC) is acknowledged for computing resources within the project jyy2580 of T.~Lappi.




\bibliographystyle{elsarticle-num} 
\bibliography{doubleD}



\end{document}